\documentclass[english,aps,manuscript]{revtex4}
\usepackage[T1]{fontenc}
\usepackage[latin9]{inputenc}
\usepackage{color}
\usepackage{esint}

\makeatletter
\@ifundefined{textcolor}{}
{%
 \definecolor{BLACK}{gray}{0}
 \definecolor{WHITE}{gray}{1}
 \definecolor{RED}{rgb}{1,0,0}
 \definecolor{GREEN}{rgb}{0,1,0}
 \definecolor{BLUE}{rgb}{0,0,1}
 \definecolor{CYAN}{cmyk}{1,0,0,0}
 \definecolor{MAGENTA}{cmyk}{0,1,0,0}
 \definecolor{YELLOW}{cmyk}{0,0,1,0}
}

\usepackage{babel}

\makeatother

\usepackage{babel}
\begin{document}

\title{On the kinetic foundations of Kaluza's magnetohydrodynamics}

\author{A. Sandoval-Villalbazo$^{1}$ , A. R. Sagaceta-Mejia$^{1}$, and
A. L. Garcia-Perciante$^{2}$}

\address{$^{1}$ Depto. de Fisica y Matematicas, Universidad Iberoamericana,
Prolongacion Paseo de la Reforma 880, Mexico D. F. 01219, Mexico.
\\
 $^{2}$ Departamento de Matematicas Aplicadas y Sistemas, Universidad
Autonoma Metropolitana-Cuajimalpa. \\
 Av. Vasco de Quiroga 487, Mexico, D.F. 05300, Mexico.}
\begin{abstract}
Recent work has shown the existence of a relativistic effect present
in a single component non-equilibrium fluid, corresponding to a heat
flux due to an electric field \cite{benedicks}. The treatment in
that work was limited to a four-dimensional Minkowksi space-time in
which the Boltzmann equation was treated in a special relativistic
approach. The more complete framework of general relativity can be
introduced to kinetic theory in order to describe transport processes
associated to electromagnetic fields. In this context the original
Kaluza's formalism is a promising approach \cite{kaluza,hyperbolic,pop}.
The present work contains a kinetic theory basis for Kaluza's magnetohydrodynamics
and gives a novel description for the establishment of thermodynamic
forces beyond the special relativistic description. 
\end{abstract}
\maketitle

\section{Introduction }

Transport processes in fluids are describable in terms of statistical
averages of microscopic collisional invariants which evolve according
to macroscopic conservation laws. The mathematical form of the corresponding
evolution equations is based on the knowledge of a distribution function
and a master equation which rules its space-time behavior. Thermodynamic
variables such as internal energy, heat flux and hydrodynamic velocity
are identified with moments of the distribution and their expressions
satisfy a set of partial differential equations that depend on the
regime of validity of the distribution function. In the case of dilute
gases, the Boltzmann equation has successfully described transport
processes within the Chapman-Enskog expansion of the distribution
function in terms of powers of the Knudsen number \cite{cc}. 

Nevertheless, when the gas is under the \emph{action of a field},
the fluxes caused by the gradients inherent to it may be treated in
the kinetic theory formalism in more than one manner. In the standard
classical formalism, the individual particle's acceleration is coupled
to the fields through Newton's law $\dot{v}^{\ell}=F^{\ell}/m$. In
this case, the Boltzmann equation for a single component system reads
\begin{equation}
\frac{\partial f}{\partial t}+v^{\ell}\frac{\partial f}{\partial x^{\ell}}+\frac{F^{\ell}}{m}\frac{\partial f}{\partial v^{\ell}}=J\left(ff'\right).\label{eq:0}
\end{equation}

Equation (\ref{eq:0}) is an evolution equation for the distribution
function $f\left(x^{\ell},v^{\ell},t\right)$ which is the density
of gas particles in phase space. Here $\ell=1,2,3$ and Einstein's
convention for equal indices has been used as in the rest of the work.
The components of the external force are given by $F^{\ell}$ and
$m$ is the mass of the individual molecules. The term of the right
hand side includes the changes in the number of particles in each
cell of phase space due to collisions and its argument $ff'$ that
it is a function of the product of the distribution functions before
and after the interactions. 

If the force included in Eq. (\ref{eq:0}) is conservative, $F^{\ell}$
can be written as the gradient of a scalar field $\psi$ such that
$\dot{v}^{\ell}\propto\psi^{,\ell}$ (the comma denotes spatial derivatives).
Thus the spatial inhomogeneities of the field constitute, in principle,
thermodynamic forces capable of creating non-equilibrium fluxes. These
thermodynamic forces are responsible of well-known non-equilibrium
effects (Benedicks, Thomson, etc). It is interesting to notice that
in the case of a single component system those fluxes vanish when
the Chapman-Enskog expansion is used to solve Boltzmann's equation
in the Navier-Stokes regime. This fact will be reviewed in section
2 of this work.

This problem is partially overcome if the first term of Eq. (\ref{eq:0})
is arbitrarily suppressed \cite{mc kelvey,spitzer} or if the fluxes
are established for a binary mixture in the Lorentz gas case \cite{cc,kremerohm}.
Simple effects such as electric conduction and thermoelectricity require
one of those approaches in order to describe experimental results
for monocomponent charged dilute gases.

In this work we propose the use of a different treatment of the Boltzmann
equation in the spirit of the general theory of relativity. In this
approach, individual particles ``fall freely'' in space-time so
that curvature constitutes a thermodynamic force in addition to local
variables' gradients ($\nabla T$, $\nabla n$, $\nabla\vec{u}$).
In this type of formalism the term $\dot{v}^{\ell}\frac{\partial f}{\partial v^{\ell}}$
of Eq. (\ref{eq:0}) can be suppressed for structureless particles
and the effect of ``external forces'' becomes present in the covariant
derivatives that are necessary to guarantee the covariance of the
transport system.

Although it is well known how to describe the motion of freely falling
particles in terms of geodesics in the case of gravitational fields
($\frac{dv^{\ell}}{dt}=-\Gamma_{ab}^{\ell}v^{a}v^{b}$), the introduction
of the concept of ``free-fall'' in the case of an electric field
is rather subtle. One elegant way to establish the equation of motion
of a simple particle imbedded in an electric field in terms of geodesics
was proposed by T. Kaluza back in 1921 \cite{kaluza}. Kaluza's idea
of a five-dimensional space-time allows to obtain Maxwell's equations
using Einstein's field equations and to reproduce the Lorentz force
exerted on a charged particle in terms of a geodesic on a curved space-time
in the presence of source charges and currents. Decades later, the
classical Kaluza framework was introduced in a phenomenological attempt
to rewrite the equations of magnetohydrodyamics \cite{pop}. The present
work provides a kinetic theory support of Kaluza's macroscopic hydrodynamics
in the presence of an electric field, yielding also a novel description
of old effects relating forces and fluxes present in the transport
theory of monocomponent dilute gases.

We have structured this paper as follows. In Sect. II we review the
standard treatment of external forces for monocomponent dilute gases
emphasizing how external forces effects are canceled in the first
order in the gradients correction of the distribution function (Navier-Stokes
regime). Also the usual argument regarding electrical conduction within
kinetic theory in the stationary approximation is critically examined.
In Sect. III we establish the system of relativistic transport equations
for a gas in a five dimensional space-time. In Sect. IV we establish
the distribution function to first order in the Knudsen parameter
following the Chapman-Enskog method and calculate of the electrical
conductivity coefficient in Kaluza's formalism. Our final remarks
are included in Sect. VI.

\section{the standard kinetic theory approach}

In this section we show how standard kinetic theory to first order
in the gradients, yields no electrostatic contribution to the non-equilibrium
distribution function for the \emph{single component} charged gas.
The non-relativistic Boltzmann equation for such a system in the presence
of an electrostatic potential $\phi$ is given by
\begin{equation}
\frac{\partial f}{\partial t}+\vec{v}\cdot\frac{\partial f}{\partial\vec{r}}+\frac{e}{m}\nabla\phi\cdot\frac{\partial f}{\partial\vec{v}}=-\frac{f-f^{\left(0\right)}}{\tau},\label{eq:01}
\end{equation}
where a relaxation time approximation has been introduced on the right
hand side. Here $f^{\left(0\right)}$ is the local equilibrium solution,
$\vec{v}$ the molecular velocity, $\tau$ a relaxation time and $m$
and $e$ the mass and electric charge of the molecules respectively.
For the first order in the gradients deviation from equilibrium $f^{\left(1\right)}=f-f^{\left(0\right)}$
one obtains
\begin{equation}
f^{\left(1\right)}=-\tau\left(\frac{\partial f^{\left(0\right)}}{\partial t}+\vec{v}\cdot\frac{\partial f^{\left(0\right)}}{\partial\vec{r}}+\frac{e}{m}\nabla\phi\cdot\frac{\partial f^{\left(0\right)}}{\partial\vec{v}}\right),\label{02}
\end{equation}
where $f^{\left(0\right)}$ is given by Maxwell-Boltzmann's distribution
function. The electric field contribution to $f^{\left(1\right)}$
arises from two terms, namely the last term on the right hand side
of Eq. (\ref{02}) and the one that appears in the time derivative
since
\begin{equation}
\frac{\partial f^{\left(0\right)}}{\partial t}=\frac{\partial f^{\left(0\right)}}{\partial n}\frac{\partial n}{\partial t}+\frac{\partial f^{\left(0\right)}}{\partial T}\frac{\partial T}{\partial t}+\frac{\partial f^{\left(0\right)}}{\partial\vec{u}}\cdot\frac{\partial\vec{u}}{\partial t}.\label{03}
\end{equation}
According to the Chapman-Enskog method, $\frac{\partial\vec{u}}{\partial t}$
can be expressed in terms of spatial gradients and external forces
in the Euler regime. Calculating explicitly both contributions one
obtains
\begin{equation}
\frac{e}{m}\nabla\phi\cdot\frac{\partial f^{\left(0\right)}}{\partial\vec{v}}=-\frac{e}{kT}f^{\left(0\right)}\nabla\phi\cdot\vec{k},\label{04}
\end{equation}
and
\begin{equation}
\frac{\partial f^{\left(0\right)}}{\partial\vec{u}}\cdot\frac{\partial\vec{u}}{\partial t}=\left(\frac{m}{kT}f^{\left(0\right)}\vec{k}\right)\cdot\left(\frac{e}{m}\nabla\phi-\nabla p\right),\label{05}
\end{equation}
where $\vec{k}=\vec{v}-\vec{u}$ is the usual peculiar velocity, $\vec{u}$
being the hydrodynamic (bulk) velocity of the fluid. From Eqs. (\ref{04})
and (\ref{05}) it is clear that this standard procedure yields no
electrostatic contribution to $f^{\left(1\right)}$, the reason being
that the drag due to the electric force on the individual particles
and on the fluid as a whole cancel each other out. This does not occur
in the mixture case.

Alternatives to obtain a dissipative electrical current for the simple
charged fluid within kinetic theory have been put forward, for example
in Refs. \cite{mc kelvey} and \cite{spitzer}. Both authors consider
a steady state case by neglecting the time derivative term in Eq.
(\ref{eq:01}). However it is important to notice that once the Chapman-Enskog
hypothesis is introduced, this term is of the same order in the Knudsen
parameter as the rest of the terms in Eq. (\ref{02}). Indeed, Eqs.
(\ref{03}) to (\ref{05}) show that this term is of first order in
the gradients as the other two terms on the right hand side of Eq.
(\ref{eq:01}) are. In the following sections of this work, a general
relativistic formalism is developed in order to establish a constitutive
equation for the electrical current using kinetic theory to first
order in the gradients without introducing approximations in the Boltzmann
equations foreign to the Chapman-Enskog scheme.

\section{Transport equations in kaluza's framework}

Kaluza's original work \cite{kaluza} establishes the possibility
of including electromagnetic effects in the geometry of space-time
by showing that a charged particle moving along a geodesic in the
proposed five-dimensional metric follows the same trajectory as the
one dictated by the Lorentz force. Such a condition is satisfied,
to first order in the components of the electromagnetic four potential,
if the metric is given by
\begin{equation}
ds^{2}=\eta_{\mu\nu}dx^{\mu}dx^{\nu}+\xi\hat{A}_{\mu}dx^{\mu}dx^{5}+\left(dx^{5}\right)^{2},\label{eq:1}
\end{equation}
where $\mu$ and $\nu$ run from 1 to 4 and $\eta_{\mu\nu}$ is the
cartesian Minkowski metric tensor. The additional elements of the
metric tensor are given by a normalized electromagnetic potential
$\hat{A}^{\mu}=\xi A^{\mu}$ where $A^{\mu}=\left[\vec{A},\,\phi/c\right]$,
with $\vec{A}$ and $\phi$ being the usual vector and scalar potential
and $\xi=\sqrt{\frac{16\pi G\epsilon_{0}}{c^{2}}}$ \cite{hyperbolic}.
In this scenario, the non-vanishing Christoffel symbols are
\begin{equation}
\Gamma_{\nu5}^{\mu}=\Gamma_{5\nu}^{\mu}=\frac{\xi}{2}F_{\nu}^{\mu},\label{eq:2}
\end{equation}
and
\begin{equation}
\Gamma_{\nu4}^{5}=-\frac{\xi}{2}F_{\nu}^{\mu},\label{eq:3-1}
\end{equation}
where $F_{\mu\nu}=A_{\mu;\nu}-A_{\nu;\mu}=A_{\mu,\nu}-A_{\nu,\mu}$
is Faraday's field tensor. 

The equation of motion for a particle in the space-time given by Eq.
(\ref{eq:1}) is determined by the geodesic equation
\begin{equation}
\frac{dv^{\mu}}{d\tau}+\Gamma_{\alpha\lambda}^{\mu}v^{\alpha}v^{\lambda}=0,\label{eq:3}
\end{equation}
where $v^{\mu}=\frac{dx^{\mu}}{d\tau}$ and $\tau$ is the particle's
proper time. Comparison of Eq. (\ref{eq:3-1}) with the equation of
motion for a charged particle in the presence of an electromagnetic
field $F^{\mu\nu}$
\begin{equation}
\frac{dv^{\mu}}{d\tau}=-\frac{e}{m}v_{\alpha}F^{\mu\alpha},\label{eq:a}
\end{equation}
gives rise to the relation proposed by Kaluza: $v^{5}=\frac{q}{\xi m}$.
Also, the so-called cylindrical condition that all derivatives with
respect to $x^{5}$ vanish is imposed in the original formalism. This
condition leads to the concept of periodic, compact extra dimensions
\cite{klein}.

For the kinetic theory purposes, the charged fluid's state variables
are given as 
\begin{equation}
\left\langle \psi\right\rangle =\frac{1}{n}\int\psi fd^{*}v,\label{eq:5}
\end{equation}
where $\psi$ is taken as $nv^{\mu}$ for the particle four-flux ($n$
being the number density) and as $mv^{\mu}v^{\nu}$ for the energy-momentum
tensor, for details on the nature of these quantities the reader can
refer to Ref. \cite{microscopic}. The average is performed over the
three dimensional velocity space, the corresponding invariant volume
element is given, as in the special relativistic case, by
\begin{equation}
d^{*}v=\gamma_{\left(w\right)}^{5}c\frac{d^{3}w}{v^{4}}=\gamma_{\left(w\right)}^{4}d^{3}w,\label{6}
\end{equation}
where $w^{\ell}$ is the vector three-velocity and $\gamma_{\left(w\right)}=\left(1-\frac{w^{2}}{c^{2}}\right)^{-1/2}$
the usual Lorentz factor. A simple derivation of this volume element
can be found in Appendix B of Ref. \cite{Heat}.

Balance equations for state variables can be obtained in this framework
by multiplying the relativistic Boltzmann's equation
\begin{equation}
v^{\mu}f_{,\mu}+\dot{v}^{\mu}\frac{\partial f}{\partial v^{\mu}}=J(ff'),\label{eq:6}
\end{equation}
by a function of the molecular four-velocity $\psi\left(v^{\mu}\right)$,
and integrating in velocity space. Here a comma denotes a partial
derivative and the dot corresponds to the total derivative 
\begin{equation}
\dot{A}^{\nu}=v^{\mu}A_{;\mu}^{\nu}=v^{\mu}\left(\frac{\partial A^{\nu}}{\partial x^{\mu}}+\Gamma_{\mu\lambda}^{\nu}A^{\lambda}\right),\label{9}
\end{equation}
where a semicolon is used to indicate a covariant derivative. The
distribution function $f$ corresponds to the particle density in
three dimensional phase space such that $f\left(x^{\mu},v^{\mu}\right)d^{3}xd^{3}v$
yields the number of particles in a volume $d^{3}xd^{3}v$. The collisional
kernel on the right hand side of Eq. (\ref{eq:6}) accounts for particles
entering and leaving a cell in phase space due to molecular interactions.
For the second term in the left hand side of Eq. (\ref{eq:6}) one
notices that 
\begin{equation}
\dot{v}^{\mu}=v^{\alpha}v_{;\alpha}^{\mu}=v^{\alpha}\left(\frac{\partial v^{\mu}}{\partial x^{\alpha}}+\Gamma_{\alpha\lambda}^{\mu}v^{\lambda}\right),\label{10}
\end{equation}
or
\begin{equation}
\dot{v}^{\mu}=\frac{dv^{\mu}}{d\tau}+\Gamma_{\alpha\lambda}^{\mu}v^{\lambda},\label{eq:qqq}
\end{equation}
which \emph{vanishes since particles move following the geodesic equation
given by Eq. (\ref{eq:3-1})}. This leads to the force-free Boltzmann
equation
\begin{equation}
v^{\mu}\frac{\partial f}{\partial x^{\nu}}=J(ff'),\label{eq:10}
\end{equation}
which is consistent with the present approach where the effects of
fields must be included in the geometry of space-time. 

Multiplying Eq. (\ref{eq:10}) by a collisional invariant and integrating
in velocity space leads to
\begin{equation}
\left[\int f\psi(v^{\alpha})v^{\mu}d^{*}v\right]_{;\mu}=0,\label{eq:8}
\end{equation}
where we have used the relation
\begin{equation}
\left[v^{\mu}\psi(v^{\alpha})f\right]_{;\mu}=\left[v^{\mu}\psi(v^{\alpha})\right]_{;\mu}f+\left[v^{\mu}\psi(v^{\alpha})\right]f_{,\mu}.\label{eq:9}
\end{equation}
In Eq. (\ref{eq:9}) the first term on the right hand side vanishes
since $\psi$ only depends explicitly on the molecular velocity. This
procedure leads to the usual balance equations for the conserved fluxes:
\begin{equation}
N_{;\mu}^{\mu}=0,\, T_{;\mu}^{\alpha\mu}=0,\label{eq:15-1}
\end{equation}
with $N^{\mu}$ and $T^{\alpha\mu}$ being the particle four-flux
and energy-momentum tensor respectively, given by 
\begin{equation}
N^{\mu}=\int fv^{\mu}d^{*}v,\label{eq:10-1}
\end{equation}
and 
\begin{equation}
T^{\mu\alpha}=m\int fv^{\mu}v^{\alpha}d^{*}v.\label{eq:11}
\end{equation}
In particular, in a local equilibrium situation, the conserved fluxes
are
\begin{equation}
N^{\mu}=n\mathcal{U}^{\mu},\label{eq:12-1}
\end{equation}
and
\begin{equation}
T^{\mu\nu}=\tilde{\rho}\,\mathcal{U}^{\mu}\mathcal{U}^{\nu}+pg^{\mu\nu},\label{eq:13-1}
\end{equation}
where $\mathcal{U}^{\nu}$ is the hydrodynamic four velocity and $\tilde{\rho}=\left(n\varepsilon+p\right)/c^{2}$
with $\varepsilon$ and $p$ being the internal energy and hydrostatic
pressure respectively \cite{microscopic}. In such regime Eqs. (\ref{eq:15-1})
evaluated for $f=f^{\left(0\right)}$ correspond to Euler's equations
for a charged single component fluid in the presence of an electrostatic
field. This set can be written as
\begin{equation}
\mathcal{U}^{\alpha}n_{,\alpha}=-n\mathcal{U}_{;\mu}^{\mu},\label{eq:17}
\end{equation}
 
\begin{equation}
\mathcal{U}^{\nu}T_{,\nu}=-\frac{p}{nC_{n}}\mathcal{U}_{;\nu}^{\nu},\label{eq:18}
\end{equation}
 
\begin{equation}
\tilde{\rho}\,\mathcal{U}^{\nu}\mathcal{U}_{;\nu}^{\mu}=-\left(g^{\mu\nu}+\frac{\mathcal{U}^{\mu}\mathcal{U}^{\nu}}{c^{2}}\right)p_{,\nu},\label{eq:19}
\end{equation}
where we have introduced the specific heat at constant number density
defined as $C_{n}=\left(\frac{\partial\varepsilon}{\partial T}\right)_{n}$.
In Eq. (\ref{eq:19}), which corresponds to the momentum balance equation
for $\mu=1,2,3$, the electric force acting on the fluid \emph{emerges
as a consequence of curvature} through the Christoffel symbols included
on the left hand side where
\begin{equation}
\mathcal{U}^{\nu}\mathcal{U}_{;\nu}^{\mu}=\mathcal{U^{\nu}}\frac{\partial\mathcal{U}^{\mu}}{\partial x^{\nu}}+\Gamma_{\nu\lambda}^{\mu}\mathcal{U}^{\mu}\mathcal{U}^{\lambda}.\label{eq:a-1}
\end{equation}
Equations (\ref{eq:17}-\ref{eq:19}) are thus the set of transport
equations for the charged single fluid in Kaluza's framework in the
local equilibrium, Euler regime. The momentum balance given by Eq.
(\ref{eq:19}) will be used in the next section in order to calculate
the contribution of the electrostatic field to the distribution function
and thus to the dissipative effects.

\section{The distribution function in Marle's approximation}

In this section, the Chapman-Enskog method of solution for the Boltzmann
equation is used in order to obtain the basic structure of the out
of equilibrium distribution function to first order in the gradients,
including a term dependent on the electrostatic field. In order to
accomplish this task, we start by stating the Chapman-Enskog hypothesis,
namely 
\begin{equation}
f=f^{(0)}+\epsilon f^{(1)}+\ldots,\label{eq:12}
\end{equation}
where $\epsilon$ is Knudsen's parameter which measures the degree
in which the distribution function deviates from equilibrium accounting
for the relative scale in which macroscopic gradients are significant
to the microscopic length. Thus, the first order in $\epsilon$ approximation
in Eq. (\ref{eq:12}) corresponds to the first order in the gradients
scenario, namely the Navier-Stokes regime. Here the equilibrium distribution
function $f^{\left(0\right)}$ is given by the relativistic Maxwellian
(or Juttner function) namely,
\begin{equation}
f^{(0)}=\frac{n}{4\pi c^{3}zK_{2}\left(\frac{1}{z}\right)}e^{\frac{\mathcal{U}_{\beta}v^{\beta}}{zc^{2}}},\label{eq:13}
\end{equation}
where $\beta=1,2,3,4$. In Eq. (\ref{eq:13}) $K_{2}$ is the modified
Bessel function of the second kind and $z=\frac{kT}{mc^{2}}$ is the
relativistic parameter that compares the thermal energy with the rest
energy of a single particle of mass $m$.

Moreover, in order to establish the structure of $f^{\left(1\right)}$
in terms of the gradients, a relaxation time approximation is appropriate
since the flux-force relations in the case of conservative forces
is retained in such model. The complete Kernel would only add precision
to the transport coefficients which is beyond the task at hand. Thus,
for the sake of clarity and simplicity, we consider Marle's approximation
and write 
\begin{equation}
J(ff^{'})=-\frac{f-f^{(0)}}{\tau},\label{eq:kerne}
\end{equation}
where $\tau$ is the relaxation parameter which scales as the microscopic
characteristic time. Introducing Eq. (\ref{eq:12-1}) in Boltzmann's
equation with the approximation given by Eq. (\ref{eq:kerne}) leads
to

\begin{equation}
\epsilon f^{(1)}=-\tau v^{\mu}\left[\frac{\partial f^{(0)}}{\partial n}n_{,\mu}+\frac{\partial f^{(0)}}{\partial T}T_{,\mu}+\frac{\partial f^{(0)}}{\partial u^{\alpha}}\mathcal{U}_{;\mu}^{\alpha}\right],\label{eq:f1}
\end{equation}
where the functional hypothesis, stating that the distribution function
depends on space and time only through the state variables, has been
also used. The derivatives of the distribution function can be readily
obtained and are given by
\begin{equation}
\frac{\partial f^{\left(0\right)}}{\partial n}=\frac{f^{\left(0\right)}}{n},\label{eq:14-1}
\end{equation}
 
\begin{equation}
\frac{\partial f^{\left(0\right)}}{\partial T}=\frac{f^{\left(0\right)}}{T}\left(1+\frac{\mathcal{U}^{\beta}v_{\beta}}{zc^{2}}-\frac{\mathcal{G}\left(\frac{1}{z}\right)}{z}\right),\label{eq:14.5-1}
\end{equation}
and
\begin{equation}
\frac{\partial f^{\left(0\right)}}{\partial\mathcal{U}^{\alpha}}=\frac{v_{\alpha}}{zc^{2}}f^{\left(0\right)},\label{eq:15}
\end{equation}
which, when introduced in Eq. (\ref{eq:f1}) lead to
\begin{equation}
\epsilon f^{(1)}=-\tau v^{\mu}f^{\left(0\right)}\left[\frac{1}{n}n_{,\mu}+\left(1-\frac{v^{\nu}\mathcal{U}_{\nu}}{zc^{2}}-\frac{\mathcal{G}\left(\frac{1}{z}\right)}{z}\right)\frac{1}{T}T_{,\mu}+\frac{v_{\alpha}}{zc^{2}}\mathcal{U}_{;\mu}^{\alpha}\right],\label{eq:ad}
\end{equation}
where $\mathcal{G}\left(\frac{1}{z}\right)=K_{3}\left(\frac{1}{z}\right)/K_{2}\left(\frac{1}{z}\right)$.

Dissipative fluxes arise from the out of equilibrium part of the distribution
function $f^{\left(1\right)}$ and are coupled linearly, in the Navier-Stokes
regime, with the thermodynamic forces of the same tensorial rank.
These forces are given by the gradients of the state variables. In
particular, the Chapman-Enskog procedure demands, for existence of
the solution, imposing the lower order balance equations as a way
of writing the state variables' time derivatives, contained in the
four-gradients in Eq. (\ref{eq:ad}), in terms of the corresponding
spatial gradients. Thus, in the Navier-Stokes regime, the Euler local
equilibrium equations Eqs. (\ref{eq:17}-\ref{eq:19}), need to be
introduced in Eq. (\ref{eq:ad}). In order to perform such a task
it is convenient to separate time and space derivatives in each term.
Also, since $\epsilon f^{(1)}$ is an invariant, it can be calculated
in the fluid's comoving frame. The corresponding equation, which is
derived in detail in the Appendix, reads 
\begin{equation}
\epsilon f^{(1)}=-\tau\gamma_{\left(k\right)}f^{\left(0\right)}\left\{ k^{\ell}\left[\frac{n_{,\ell}}{n}+\left(1+\frac{\gamma_{\left(k\right)}}{z}-\frac{\mathcal{G}\left(\frac{1}{z}\right)}{z}\right)\frac{T_{,\ell}}{T}\right]+\frac{\gamma_{\left(k\right)}k_{\alpha}k^{\mu}}{zc^{2}}\mathcal{U}_{;\mu}^{\alpha}\right\} ,\label{as}
\end{equation}
where $\ell=1,2,3$. Here $k^{\ell}$ is the chaotic or peculiar velocity
which corresponds to the velocity of the molecules measured in the
comoving frame. 

The last term in Eq. (\ref{as}) includes the contribution of the
electromagnetic potential to the distribution function, as emphasized
in the previous section. Focusing only on that term one obtains ($\ell=1,2,3$)
\begin{equation}
k_{\alpha}k^{\mu}\mathcal{U}_{;\mu}^{\alpha}=k_{\alpha}k^{\ell}\frac{\partial\mathcal{U}^{\alpha}}{\partial x^{\ell}}+k_{\alpha}\frac{\partial\mathcal{U}^{\alpha}}{\partial\bar{t}}+\Gamma_{\mu\nu}^{\alpha}\mathcal{U}^{\nu}k^{\mu}k_{\alpha}.\label{eq:ar}
\end{equation}
Notice that this expression contains two contributions. The last term
on the right hand side of Eq. (\ref{eq:ar}) is due to the effect
of curvature on the motion of the molecules. Meanwhile, the second
term is also affected by the electromagnetic field since
\[
k_{\alpha}\frac{\partial\mathcal{U}^{\alpha}}{\partial\bar{t}}=-\frac{1}{\tilde{\rho}}\left(g^{\alpha\nu}+\frac{\mathcal{U}^{\alpha}\mathcal{U}^{\nu}}{c^{2}}\right)k_{\alpha}p_{,\nu}-\Gamma_{\nu\lambda}^{\alpha}\mathcal{U}^{\lambda}\mathcal{U}^{\nu}k_{\alpha},
\]
where Eq. (\ref{eq:19}) has been introduced. Notice that this effect
corresponds to the electromagnetic field acting, through the curvature
of space-time, on the motion of the fluid \emph{as a bulk}. These
two terms, both proportional to the components of the Faraday tensor,
are present in $\epsilon f^{\left(1\right)}$ and are thus capable
of giving rise to dissipation. In the case of a purely electrostatic
field and since all terms are calculated in the comoving frame one
obtains
\begin{equation}
\Gamma_{\nu\lambda}^{\alpha}\mathcal{U}^{\lambda}\mathcal{U}^{\nu}k_{\alpha}=2\frac{qc}{\xi m}\Gamma_{54}^{\ell}k_{\ell},\label{eq:e}
\end{equation}
and
\begin{equation}
\Gamma_{\nu\lambda}^{\alpha}\mathcal{U}^{\lambda}k^{\nu}k_{\alpha}=0.\label{eq:b}
\end{equation}
Introducing these results in Eq. (\ref{as}) and defining as $\epsilon f_{E}^{(1)}$
the part of the distribution function depending on the electric field
one obtains

\begin{equation}
\epsilon f_{E}^{(1)}=\frac{2q}{\xi mc}\frac{\tau}{z}\gamma_{\left(k\right)}^{2}f^{\left(0\right)}\Gamma_{54}^{\ell}k_{\ell},\label{as-1}
\end{equation}
which, using Eq. (\ref{eq:2}) yields
\begin{equation}
\epsilon f_{E}^{(1)}=\tau\frac{q}{kT}f^{\left(0\right)}\gamma_{\left(k\right)}^{2}k_{\ell}\phi^{,\ell}.\label{eq:s}
\end{equation}
Equation ($\ref{eq:s}$) shows that in Kaluza's framework the non-equilibrium
part of the distribution function features a contribution arising
from the electric field which leads to dissipative fluxes in the single
component fluid. As mentioned in Sect. I, the establishment of an
electrical conductivity linked to the electrostatic potential through
Ohm's laws is of particular interest.

The general expression for the electrical current in relativistic
kinetic theory is given by the average \cite{microscopic}

\begin{equation}
J^{\beta}=\int\left(q\gamma_{\left(k\right)}k^{\beta}\right)\epsilon f^{\left(1\right)}d^{*}K.\label{eq:heatFlux rel-1-2}
\end{equation}
The integral can be readily evaluated using

\begin{equation}
d^{*}K=4\pi c^{3}\sqrt{\gamma_{\left(k\right)}^{2}-1}d\gamma,\label{eq:44}
\end{equation}
and
\begin{equation}
k^{2}=c^{2}\left(\frac{\gamma_{\left(k\right)}^{2}-1}{\gamma_{\left(k\right)}^{2}}\right).\label{85}
\end{equation}
Using the previous equations one obtains the following expression
for the electrical current
\begin{equation}
J^{\ell}=\tau\frac{nq^{2}}{m}\mathcal{G}\left(\frac{1}{z}\right)\phi^{,\ell}.\label{55}
\end{equation}
The heat flux in Eq. (\ref{55}) can be written in the low $z$ limit
as

\begin{equation}
J^{\ell}\sim\tau\frac{nq^{2}}{m}\left(1+\frac{5}{2}z+...\right)\phi^{,\ell},\label{33}
\end{equation}
which is consistent with the non-relativistic results obtained in
Ref. \cite{mc kelvey,spitzer}.

\section{Final remarks}

The concept of external forces is foreign to the spirit of the theory
of general relativity, external forces are replaced by space-time
curvature. Following this idea, it is natural to explore the possibility
of introducing space-time curvature into kinetic theory suppressing
the external force term acting on single particles in Boltzmann's
equation. While the introduction of curvature in the gravitational
field case is rather straightforward, the electromagnetic counterpart
has been an open question since the introduction of general relativity.
One approach that has proved useful in describing the dynamics of
a charged particle in space time in terms of geodesics is the Kaluza's
original formalism first introduced in 1921. 

In this work we show how the old problem regarding how to obtain Ohm's
law for a \emph{single component} charged dilute gas from kinetic
theory consistently with the Chapman-Enskog expansion is solved if
Kaluza's description is applied to Boltzmann's equation. Moreover,
the establishment of vector fluxes such as heat and the study of thermoelectric
effects through these techniques is feasible and will be addressed
in the near future.

In a recent publication two of us calculated the counterpart of Eq.
(\ref{as-1}) in special relativity introducing the gradient of the
electrostatic potential as an external force \cite{benedicks}. In
that work, the deviation from equilibrium due to this force vanishes
in the non-relativistic limit. In that case, the electrostatic contribution
to lowest order in $z$ cancels due to the same reasons as the ones
exhibited in Sec. II, that is the equivalency of the force acting
on the fluid as a whole and on the individual molecules. On the other
hand, as here shown, Ohm's law can be obtained for a single species
charged fluid identifying the electrostatic force as curvature. In
this case the molecules move on geodesics and thus the acceleration
term in Boltzmann's equation is not present, leaving the geometry
to act on the distribution function solely through the momentum balance
equation for the fluid.

As a final comment we wish to point out that the fact that a problem
related with the electrical conductivity for a single component \emph{non-relativistic}
gas has been solved here using a relativistic formalism is somehow
natural since all electromagnetic phenomena possess a relativistic
character. Other relativistic and magnetic effects will also be accounted
for in future work.

\section*{Appendix}

In this appendix, the steps leading to Eq. (\ref{as}) are shown to
some detail. After Chapman-Enskog's hypothesis has been introduced
in Boltzmann's equation together with the derivatives of Juttner's
distribution function, one obtains Eq. (\ref{eq:ad}) in which time
and space derivatives need to be separated. Chapman-Enskog's method
requires the introduction of Euler's equations in order to write time
derivatives in terms of spatial gradients in the Navier-Stokes regime. 

For the sake of clarity, such separation is performed only for the
density and temperature derivatives as a first step in the main paper.
In this way one can isolate the term corresponding to the derivative
of the hydrodynamic velocity from which the electric contributions
will arise. Also, since the calculation is carried out in the comoving
frame one has $\mathcal{U}^{\alpha}=\left[\vec{0},c\right]$ and $v^{\alpha}=\gamma_{\left(k\right)}\left[\vec{k},c\right]$,
where $\vec{k}$ is the velocity measured in such frame, that is the
chaotic or peculiar velocity \cite{microscopic}. Proceeding in such
a way one can rewrite Eq. \textcolor{blue}{(\ref{eq:ad})} as follows
\begin{eqnarray}
\varepsilon f^{(1)}=-\tau f^{\left(0\right)}\left\{ \gamma_{\left(k\right)}\left[\frac{1}{n}\frac{\partial n}{\partial t}+\left(1-\frac{v^{\nu}\mathcal{U}_{\nu}}{zc^{2}}-\frac{\mathcal{G}\left(\frac{1}{z}\right)}{z}\right)\frac{1}{T}\frac{\partial T}{\partial t}\right]-\right.\nonumber \\
\left.-\gamma_{\left(k\right)}k^{\ell}\left[\frac{1}{n}n_{,\ell}+\left(1-\frac{v^{\nu}\mathcal{U}_{\nu}}{zc^{2}}-\frac{\mathcal{G}\left(\frac{1}{z}\right)}{z}\right)\frac{1}{T}T_{,\ell}\right]-\frac{\gamma_{\left(k\right)}^{2}k^{\mu}k_{\alpha}}{zc^{2}}\mathcal{U}_{;\mu}^{\alpha}\right\} \label{eq:ad-1}
\end{eqnarray}
where now $\ell=1,2,3$ and the greek indices $\mu$ and $\nu$ still
run up to $4$. 

As mentioned above and in the main text, time derivatives need to
be written in terms of thermodynamic forces by means of Euler's equations
which can be readily obtained from Boltzmann's equation. Indeed, for
the density one uses the continuity equation, Eq. (\ref{eq:17}),
which in the comoving frame yields
\[
\frac{\partial n}{\partial t}=0
\]
For the temperature one uses the internal energy balance together
with $\varepsilon_{,\nu}=C_{n}T_{,\nu}$ which yields Eq. (\ref{eq:18}).
Thus in the comoving frame one also has that 
\[
\frac{\partial T}{\partial t}=0
\]
Because of this, the first square bracket in Eq. (\ref{eq:ad-1})
vanishes so that 
\begin{equation}
\varepsilon f^{(1)}=\tau f^{\left(0\right)}\gamma_{\left(k\right)}\left\{ k^{\ell}\left[\frac{1}{n}n_{,\ell}+\left(1-\frac{v^{\nu}\mathcal{U}_{\nu}}{zc^{2}}-\frac{\mathcal{G}\left(\frac{1}{z}\right)}{z}\right)\frac{1}{T}T_{,\ell}\right]-\frac{\gamma_{\left(k\right)}k^{\mu}k_{\alpha}}{zc^{2}}\mathcal{U}_{;\mu}^{\alpha}\right\} \label{eq:ad-1-1}
\end{equation}
which is precisely Eq. (\ref{as}) in the main text introducing $v^{\nu}\mathcal{U}_{\nu}=-\gamma_{\left(k\right)}c^{2}$.
The subsequent separation and careful treatment of the last term,
which involves the covariant derivative, is carried out to detail
in Eqs. (\ref{eq:ar}-\ref{eq:s}).

\textsf{\textbf{\textit{\Large Acknowledgements}}}{\Large \par}

The authors acknowledge support from CONACyT through grant number
CB2011/167563. The authors also wish to thank D. Brun-Battistini for
her valuable comments for the final version of this work.


\begin{thebibliography}{10}
\bibitem[1]{benedicks} A.L. Garcia-Perciante, A. Sandoval-Villalbazo,
L.S. Garcia-Colin, Benedicks effect in a relativistic simple fluid,
Jour. Non-Equilib. Thermodyn. \textbf{38} (2013) 141\textendash{}151.

\bibitem[2]{kaluza} T. Kaluza, Zum unitatsproblem der physik, Sitzungsbrichte
d. Berl. Akad. (1921).

\bibitem[3]{pop}A. Sandoval-Villalbazo, L. S. Garcia-Colin, Relativistic
magnetohydrodynamics revisited, Phys. Plasmas \textbf{7}, 4823 (2000)
.

\bibitem[4]{hyperbolic} A. Sandoval-Villalbazo, A. L. Garcia-Perciante,
L. S. Garcia-Colin, Hyperbolic heat equation in Kaluza's magnetohydrodynamics,
Gen. Rel. Grav. \textbf{39}, 1287 (2007).

\bibitem[5]{cc}S. Chapman, and T. G. Cowling, The mathematical theory
of non-uniform gases (Cambridge Mathematical Library, United Kingdom,
1971), 3rd. ed.

\bibitem[6]{mc kelvey}J. P. McKelvey; Solid State and Semiconductor
Physics, Krieger Pub Co., (1982).

\bibitem[7]{spitzer}Cohen, R.S., Spitzer Jr., L., McRoutty, P., The
electrical conductivity of an ionized gas, Phys. Rev., \textbf{80}
(1950), 230; Spitzer Jr., Physics of Fully Ionized Gases, Dover Publications
(2006).

\bibitem[8]{kremerohm}G. M. Kremer, C. H. Patsko, Relativistic ionized
gases: Ohm and Fourier laws from Anderson and Witting model equation,
Physica A \textbf{322, }329 (2003).

\bibitem[9]{klein}Klein O., Nature \textbf{110}, 516 (1926).

\bibitem[10]{microscopic}García-Perciante A. L., Sandoval-Villalbazo
A., García-Colín L. S, On the microscopic nature of dissipative effects
in special relativistic kinetic theory, Jour. Non-Equilib. Thermodyn.
\textbf{37}, 43 (2012).

\bibitem[11]{Heat} Garcia-Perciante A. L. and Mendez A. R., Heat
conduction in relativistic neutral gases revisited, Gen. Rel. Grav.
\textbf{43} 2257-2275, 2011.\end{thebibliography}
\end{document}